\documentclass[aps,prb,twocolumn,superscriptaddress,showpacs]{revtex4-2}
\usepackage{graphicx,bm,color,float,textcomp,mathtools,times,nicefrac}
\newcommand\sho{$\rm SrHo_2O_4$}
\newcommand\sgo{$\rm SrGd_2O_4$}
\newcommand\seo{$\rm SrEr_2O_4$}
\newcommand\sdo{$\rm SrDy_2O_4$}
\newcommand\syo{$\rm SrYb_2O_4$}
\newcommand\sno{$\rm SrNd_2O_4$}
\newcommand\alo{$ALn_2{\rm O}_4$}
\newcommand\slo{${\rm Sr}Ln_2{\rm O}_4$}
\newcommand\bno{$\rm BaNd_2O_4$}
\newcommand\bdo{$\rm BaDy_2O_4$}
\newcommand\bto{$\rm BaTb_2O_4$}

\newcommand\afm{antiferromagnetic}
\newcommand\QV{$\mathbf{q} = (0 \frac{1}{2} \frac{1}{2})$}

\usepackage{soul,xcolor}
\setstcolor{magenta}

\begin{document}
	\title{Magnetic structure and low temperature properties of geometrically frustrated SrNd$_2$O$_4$}
\date{\today}
\author{N.~Qureshi}			\affiliation{Institut Laue-Langevin, 71 Avenue des Martyrs, CS 20156, 38042 Grenoble Cedex 9, France}
\author{A.~Wildes}			\affiliation{Institut Laue-Langevin, 71 Avenue des Martyrs, CS 20156, 38042 Grenoble Cedex 9, France}
\author{C.~Ritter}			\affiliation{Institut Laue-Langevin, 71 Avenue des Martyrs, CS 20156, 38042 Grenoble Cedex 9, France}
\author{B.~F\aa k}			\affiliation{Institut Laue-Langevin, 71 Avenue des Martyrs, CS 20156, 38042 Grenoble Cedex 9, France}
\author{S.X.M.~Riberolles}	\affiliation{Department of Physics, University of Warwick, Coventry CV4 7AL, United Kingdom}
						\affiliation{Institut Laue-Langevin, 71 Avenue des Martyrs, CS 20156, 38042 Grenoble Cedex 9, France}
\author{M.~Ciomaga Hatnean}	\affiliation{Department of Physics, University of Warwick, Coventry CV4 7AL, United Kingdom}
\author{O.~A.~Petrenko}		\affiliation{Department of Physics, University of Warwick, Coventry CV4 7AL, United Kingdom}

\begin{abstract}	
We report the low-temperature properties of \sno, a geometrically frustrated  magnet.
Magnetisation and heat capacity measurements performed on polycrystalline samples indicate the appearance of a magnetically ordered state at $T_{\rm N}=2.28(4)$~K.
Powder neutron diffraction measurements reveal that an \afm\ state with the propagation vector \QV\ is stabilised below this temperature.
The magnetic order is incomplete, as only one of the two Nd$^{3+}$ sites carries a significant magnetic moment while the other site remains largely disordered.
The presence of a disordered magnetic component below  $T_{\rm N}$ is confirmed with polarised neutron diffraction measurements.
In an applied magnetic field, the bulk properties measurements indicate a phase transition at about 30~kOe.
We construct a tentative $H$-$T$ phase diagram of \sno\ from these measurements.
\end{abstract}

\maketitle
\section{Introduction}

The magnetic properties of a family of strontium rare-earth oxides with a general formula \alo\ ($A$ = Ba, Sr; $Ln$ = lanthanide) have recently attracted considerable attention, as the magnetic $Ln$ ions in these compounds form zig-zag ladders consisting of the chains of edge-sharing triangles in a honeycomb-like arrangement~\cite{Karunadasa_2005}.
This geometry, combined with the \afm\ exchange interactions results in a behaviour typically associated with geometrical frustration, such as suppression of the magnetic ordering down to the very low temperatures, complex and fragile ground states sensitive to minute perturbations as well as the appearance of magnetisation plateaux in an applied field.
Previous studies of the \slo\ family mostly focused on the heavy rare-earth members, such as \sgo~\cite{Young_2014,Hasan_2017,Jiang_2018}, \sdo~\cite{Cheffings_2013,Bidaud_2016,Petrenko_2017,Gauthier_2017a,Gauthier_2017b,Gauthier_2017c}, \sho~\cite{Young_2012,Young_2013,Fennell_2014,Wen_2015,Young_2019}, \seo~\cite{Petrenko_2008,Hayes_2011,Hayes_2012,Fennell_2014,Malkin_2015,Qureshi_2020}, and \syo~\cite{Quintero_2012}.
The magnetic properties of \sno\ containing a lighter lanthanide, Nd, have not yet been reported, although \bno\ and \bto\ have been investigated with magnetisation, heat-capacity, and powder neutron diffraction measurements~\cite{Aczel_2014,Aczel_2015}.

\begin{figure*}[tb] 
\includegraphics[width=0.81\textwidth]{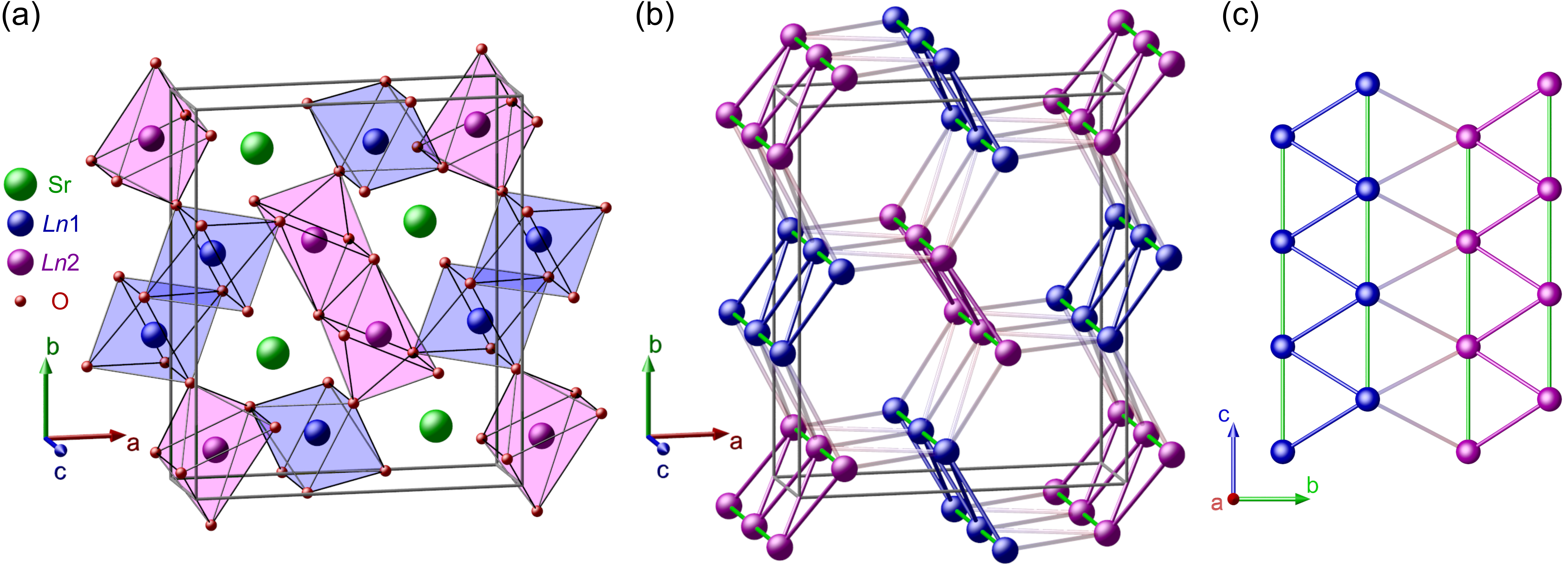}
\caption{(a) Schematic of the crystallographic unit cell of \slo\ as seen within the $a$-$b$ plane.
	The oxygen octahedra hosting both $Ln$1 and $Ln$2 ions and forming the distorted hexagonal lattice within the $a$-$b$ plane are displayed as transparent for clarity.
	(b) Three dimensional schematic of the \slo\ magnetic sublattice.
	The NN, $Ln$1 NNN and $Ln$2 NNN bonds are displayed as green, blue and purple lines, respectively.
	The inter-site bonds are shown with transparent lines and a blue-to-purple gradient.
	(c) View along the $a$ axis onto the two different types of triangular ladders.}
\label{structure}
\end{figure*}
		
The \slo\ systems have the same structure as calcium ferrite, CaFe$_2$O$_4$~\cite{Decker_1957} described by the orthorombic space group $Pnam$ in which every ion occupies a $4c$ Wyckoff position.
The unit cell contains 8 $Ln^{3+}$ magnetic ions confined within two types of oxygen octahedra with unequal degrees of distortion, see Fig.~\ref{structure}(a).
Two distinct crystallographic environments are thus present for the $Ln^{3+}$ magnetic ions which are referred to as sites $Ln$1 and $Ln$2 (shown in blue and purple, respectively, in Fig.~\ref{structure}).

The shortest distance between the $Ln^{3+}$ ions, identical for both sites, is along the $c$~axis, and in case of Nd, it is about 3.57~\AA\ (the length of the unit cell).
The  nearest neighbour (NN) bonds are shown in green in Figs.~\ref{structure}(b) and~\ref{structure}(c).
Close to the crystallographic $a$-$b$~plane, the $Ln$ ions link to their next nearest neighbours (NNN) from the same site, separated by 3.64 and 3.71~\AA\ for Nd1 and Nd2 ions shown in Figs.~\ref{structure}(b) and~\ref{structure}(c) as blue and purple bonds respectively.
This results in the formation of triangular (zig-zag) ladders propagating along the $c$~axis.
Further neighbour interactions between different $Ln$ sites link the ladders and form planes of distorted hexagons when viewed along the $c$~axis.
The inter-site bonds are significantly longer, 3.88 and 4.13~\AA\ for the case of \sno.

From the above representation of the crystal structure it follows that a full description of the magnetic interactions in the \slo\ systems should be three-dimensional in character, however, the strongest interactions are often found within the frustrated zig-zag ladders.
Indeed, in several previously reported \slo\ magnets. the inter-site magnetic interactions are significantly weaker than the intra-ladder interactions~\cite{Petrenko_2008,Young_2013}. 
Nevertheless, the inter-site interactions cannot be completely neglected~\cite{Hasan_2017,Dublenych_2020}.

The members of the \slo\ family are reported to order at temperatures well below their corresponding Weiss temperatures, e.g. \sgo~\cite{Young_2014}, \seo~\cite{Karunadasa_2005,Petrenko_2008} and \syo~\cite{Quintero_2012}, or to remain (at least partially) disordered down to the lowest temperatures e.g. \sdo~\cite{Petrenko_2017,Gauthier_2017a,Gauthier_2017b} and \sho~\cite{Ghosh_2011,Young_2013}.
Many compounds from the \slo\ family demonstrate highly anisotropic magnetic properties at low-temperature.
It is also common for the magnetic moments on the $Ln$1 and $Ln$2 sites to behave almost independently, as the intrinsic coupling between these sites is rather limited.
In some cases, a short-range order on one $Ln$ site cohabitates with a long-range order on another.

The magnetic order stabilised on the $Ln$1 sites in \sho\ and \seo\ consists of ferromagnetic chains running along the ladders legs, two adjacent legs being paired \afm ly~\cite{Young_2013,Petrenko_2008}.
This type of magnetic order is referred to as a N\'eel or up-down-up-down (UDUD) structure reflecting the sequence of spins encountered along the ladder.
Another possibility is an up-up-down-down (UUDD) arrangement, referred to as double N\'eel phase.
The two different spin sequences are believed to result from different ratios in the strength of NN and NNN exchange interactions acting within the triangular ladders~\cite{Hayes_2011,Malkin_2015,Fennell_2014}, although the inter-ladder exchange interactions between ladders made of the same sites ($Ln$1 or $Ln$2) are also believed to be influential in establishing the magnetic order, for example, in \seo~\cite{Hayes_2011}.

Alternatively, when the anisotropy is of the easy-plane type, the N\'eel or double N\'eel phases are made of spins confined within the $a$-$b$ plane of the material.
The double N\'eel arrangement of the in-plane moments is found, for instance, in \bno \cite{Aczel_2014}.

Given the large magnetic moments of the rare-earth ions and relatively weak exchange interactions, one cannot rule out the influence of the dipole-dipole interactions on the selection of the magnetic order in the \slo\ systems.
It is therefore rather interesting to investigate a variant hosting a $Ln^{3+}$ ion bearing a smaller magnetic moment, such as \sno, for which the dipolar interactions are expected  to play a less prominent role.
The \sno\ compound remained relatively unexplored compared to other members of the family, although its synthesis has been previously reported, \cite{USSR_academy,Wongng_1995,Muller-Buschbaum_2006}.
This could be explained at least partially by the fact that \sno\ is chemically unstable at high temperatures, which complicates the synthesis process, as well as by the unavailability of large single-crystal samples.

\section{Experimental details} \label{sec_methods}
	
\sno\ was synthesised in polycrystalline form by the conventional solid state synthesis method, following a procedure similar to the one employed by Karunadasa {\it et al}.~\cite{Karunadasa_2005}.
As previously reported by Wong-Ng~\cite{Wongng_1995}, \sno\ decomposes above ($1370\pm20$)~$^{\circ}$C resulting in a mixture of different chemical phases, including Sr$_3$Nd$_4$O$_9$.
We have carried out several synthesis attempts in order to determine the optimal conditions (synthesis temperature and duration) for preparing \sno.
The best results were obtained when powders of Nd$_2$O$_3$ and SrCO$_3$ were weighed in stoichiometric amounts, mixed together and heat treated in air for 48 hours, at 1300~$^{\circ}$C. 
To ensure the appropriate composition of the final compound, the Nd$_2$O$_3$ powder was pre-annealed in air at 1000~$^{\circ}$C for 24 hours.
Powder X-ray diffraction measurements confirmed the phase purity of the \sno\ polycrystalline sample synthesised.
The material slowly decomposes when exposed to air, it was therefore stored in an inert gas atmosphere.
		
Magnetisation measurements were performed on a  sintered polycrystalline sample of \sno\ using a Quantum Design MPMS magnetometer in the temperature range 1.8 to 300~K.
The magnetic moment at lower temperatures, 0.5 to 2.0~K, was measured using an iQuantum $^3$He insert~\cite{Shirakawa_2004}. 

Specific heat was measured using a Quantum Design PPMS calorimeter equipped with a $^3$He option in the temperature range 0.4 to 20~K, in fields of up to 50~kOe.
We have measured the specific heat as a function of temperature in a constant magnetic field.
The measurements were carried out on thin pressed and sintered pellets of polycrystalline material a few mg in mass.

Powder neutron diffraction (PND) measurements were performed at the Institut Laue-Langevin (ILL), Grenoble, France.
Neutron scattering techniques were employed to investigate both the nuclear structure and low temperature magnetic phases.
A series of PND patterns were collected on the high-intensity D20 and high-resolution D2B diffractometers.
Measurements were performed on D20 from room temperature down to the milli-kelvin region using a dilution fridge cryostat and a Cu sample container.
In addition, a regular Orange cryostat and a vanadium sample container were employed on both instruments from room temperature down to 1.6~K.
Wavelengths of 2.41 and 1.59~\AA\ were used on D20 and D2B, respectively.
The data have been treated employing the Rietveld refinement analysis technique with the help of the FullProf Suite software~\cite{FULLPROF}.
	
Polarised neutron diffraction experiments were also carried out at the ILL using the D7 instrument~\cite{Stewart_2009}, a cold-neutron diffuse scattering spectrometer equipped with xyz polarisation analysis.
With the three banks of 44 $^3$He detectors each, the wavelength of 3.14~\AA\ gave a $Q$-range coverage of 0.2 to 3.8~\AA$^{-1}$.
The sample was loaded into an annular shaped thin-wall aluminium container within a standard orange cryostat with a base temperature of 1.5~K.
Standard data analysis techniques (detector efficiency normalisation from vanadium standards and polarisation efficiency calculations from an amorphous silica standard) were employed to separate the magnetic, nuclear, and
spin-incoherent scattering components.

Inelastic neutron scattering (INS) measurements were performed on the direct geometry thermal time-of-flight spectrometer PANTHER at the ILL using an incoming energy of 35~meV from the (004) reflection of a double focusing pyrolythic graphite monochromator and a chopper frequency of 267~Hz.
The powder sample of weight 5.2~g was mounted in a thin-walled cylindrical aluminium container in annular geometry inside an orange cryostat.
Measurements were performed in the paramagnetic state at $T=5$ and 150~K.
Standard data analysis was used to transform the raw data (from the highly pixelated position-sensitive detector) into the dynamic structure factor $S(Q,E)$.
	
\section{Results and discussion}
\subsection{Magnetisation and heat capacity measurements} \label{section_bulk}

\begin{figure}[tb] 
\includegraphics[width=0.9\columnwidth]{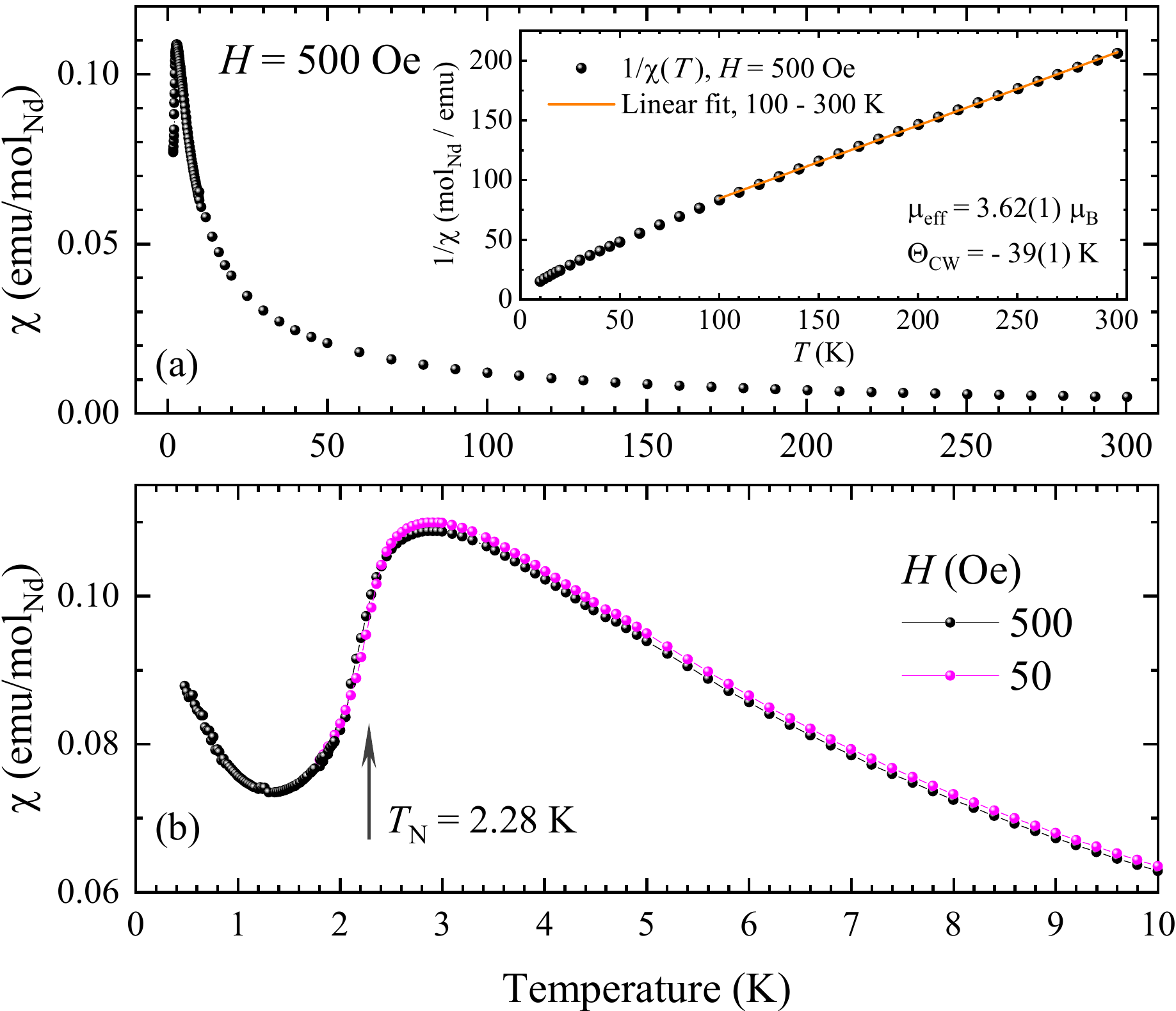}
\caption{(a) Temperature dependence of magnetic susceptibility $\chi(T)$ measured on a powder sample of \sno\ in a field of 500~Oe.
	The inset shows a linear fit to 1/$\chi(T)$ in the range 100 to 300~K.
	(b) The low-$T$ part of the $\chi(T)$ curve, which is extended down to 0.5~K for $H=500$~Oe.}
\label{Fig_MT}
\end{figure}

The temperature dependence of the magnetic susceptibility $\chi(T)$ of \sno\ is shown in Fig.~\ref{Fig_MT}.
The inverse susceptibility curve,  $\chi^{-1}(T)$, is almost linear with temperature, particularly in the higher-temperature range.
The linear fit in the range 100 to 300~K returns the value of 3.62(1)~$\mu_{\rm B}$ for an effective magnetic moment of Nd$^{3+}$ ions, $\mu_{\rm eff}$, and $\Theta_{CW}=-39(1)$~K for the Weiss temperature, suggesting a considerable strength of the \afm\ interactions present in \sno.
The obtained value of $\mu_{\rm eff}$ is exactly what would be expected for a free Nd$^{3+}$ ion, however, as the $1/\chi(T)$ curve is not perfectly linear, varying the temperature range of the fit changes the effective moment slightly.
The deviation from an ideal linear dependence is
caused by the presence of the low energy crystalline electric field (CEF) excitations (see section~\ref{section_INS}).
If the temperature range 5 to 10~K is considered (the range where the $1/\chi(T)$ curve is approximately linear and where the effects of populating the CEF levels can be neglected), the effective magnetic moment of the Nd$^{3+}$ ions decreases to 2.74(2)~$\mu_{\rm B}$.
 
At the lower temperature range, there is a broad maximum in $\chi(T)$ at approximately 2.9~K, below which the susceptibility sharply decreases.
For an applied field of 50~Oe, the susceptibility drop is the sharpest at 2.28~K [see Fig.~\ref{Fig_MT}(b)], which in combination with the results of the heat capacity measurement presented in the next section suggest a magnetic ordering temperature of $T_{\rm N}=(2.28\pm0.04)$~K.
Below $T_{\rm N}$ the susceptibility continues to decrease until about 1.4~K and then rises slightly at the lowest $T$.
We have performed the measurements with both field-cooled-warming and zero-field-cooled protocols and detected almost no difference between them apart from a very small hysteresis around $T_{\rm N}$.

\begin{figure}[tb] 
\includegraphics[width=0.9\columnwidth]{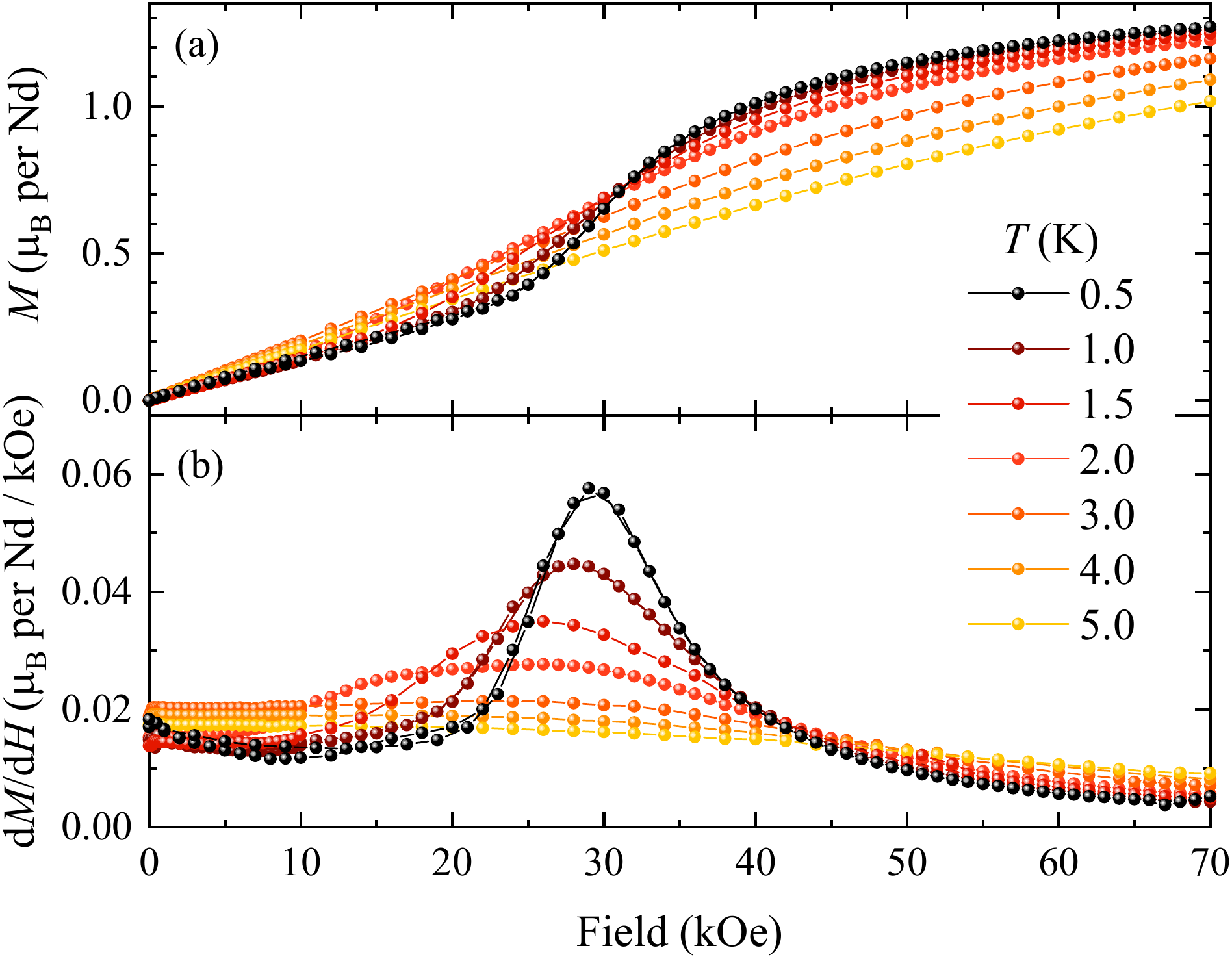}
\caption{Field dependence of (a) magnetisation $M(H)$ and (b) its field derivative $dM(H)/dH$ measured at various temperatures on a powder sample of \sno.}
\label{Fig_MH}
\end{figure}

The results of the magnetisation measurements on a sintered powder sample of \sno\ at different temperatures are shown in Fig.~\ref{Fig_MH}.
At the lowest experimentally attainable temperature of 0.5~K, the magnetisation grows linearly with field until about 20~kOe, then it shows a pronounced increase in the field range 20 to 40~kOe and a tendency to saturation above this field.
In the maximum field of 70~kOe the magnetic moment per Nd$^{3+}$ ion approaches 1.4~$\mu_{\rm B}$, which is less than half of the expected value for a free Nd$^{3+}$ ion.

For the lowest temperature $T=0.5$~K, a well-defined maximum at 29.5~kOe in the $dM(H)/dH$ curve, as shown in Fig.~\ref{Fig_MH}(b), suggests a field-induced transition.
Upon increasing the temperature, the transition shifts to slightly lower fields and becomes progressively broader.
The highest temperature at which the maximum in the $dM(H)/dH$ curve is still visible is 2.0~K while for $T=3$~K and above, the magnetisation curves are featureless and their shape resembles that of a simple paramagnet.

Overall, the magnetisation process in \sno\ looks very similar to what has been found in the isostructural compound \bno~\cite{Aczel_2014}, however, our interpretation of the transition observed in \sno\ just below 30~kOe through the magnetisation measurements is different.
While Ref.~\cite{Aczel_2014} suggests that transitions found in \bno\ are between the magnetically ordered state and the paramagnetic phase (which results in the appearance of a rather simple $H$-$T$ phase diagram depicted in Fig.~7 of Ref.~\cite{Aczel_2014}), we believe that in \sno\ the applied field first induces a spin-reorientation transition within the ordered state at around 30~kOe while magnetisation saturation takes place at higher fields, well above the 70~kOe (the maximum field available in the MPMS magnetometer).

\begin{figure}[tb] 
\includegraphics[width=0.9\columnwidth]{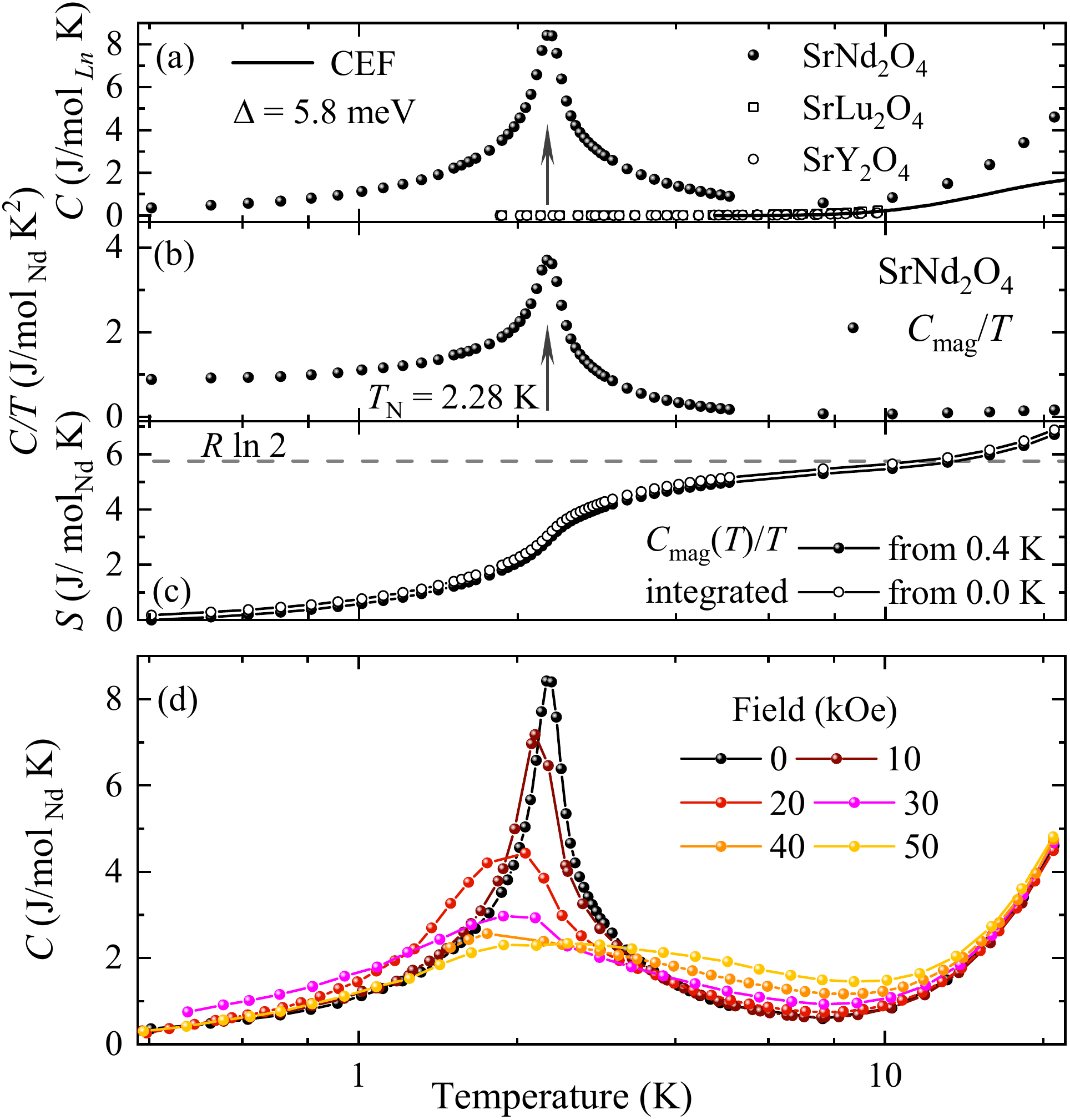}
\caption{Temperature dependence (on a log scale) of (a) heat capacity $C(T)$ of \sno\ and non-magnetic isostructural oxides {$\rm SrLu_2O_4$} and {$\rm SrY_2O_4$} in zero field;
	(b) magnetic component of the heat capacity of \sno\ divided by temperature, $C_{\rm mag}/T$, for $H=0$;
	(c) zero-field magnetic entropy, calculated as the area under the $C_{\rm mag}/T$ curve;
	(d) heat capacity of \sno\ in various applied fields.
	The solid line in (a) shows the Schottky heat capacity of a two-level system with equal degeneracy and an energy gap of $\Delta=67.3$~K.}
\label{Fig_HC}
\end{figure}

Temperature dependence of the heat capacity, $C(T)$, of \sno\ in zero field is shown in Fig.~\ref{Fig_HC}(a).
A sharp peak in $C(T)$ suggests a magnetic ordering transition at $T_{\rm N}=(2.28\pm0.04)$~K.
The top panel of Fig.~\ref{Fig_HC} also shows the heat capacity of the two non-magnetic isostructural compounds, SrLu$_2$O$_4$ and SrY$_2$O$_4$, (open symbols)~\cite{Petrenko_2008} to emphasise the fact that the lattice contribution to the heat capacity of \sno\ remains rather limited in the temperature range shown.
Therefore an increase in $C(T)$ seen in \sno\ at higher temperatures, $T>10$~K, apart from the phonon contribution must also contain a significant magnetic component caused by the low-lying CEF excitations of Nd$^{3+}$.
The solid line in Fig.~\ref{Fig_HC}(a) represents the temperature dependence of the heat capacity of a two-level system with the energy separation of 5.8~meV (67.3~K) to match the results of the INS measurements (see section~\ref{section_INS})~\cite{Fak_2020}.
The two levels were presumed to be equally degenerate for this Schottky effect calculation.

Figure~\ref{Fig_HC}(b) shows the temperature dependence of the magnetic component of the heat capacity divided by temperature, $C_{\rm mag}/T$.
$C_{\rm mag}$ was calculated by subtracting the lattice contribution (in a simple approximation that it follows the $T^3$ dependence) from the total heat capacity.
Remarkably the $C_{\rm mag}(T)/T$ seems to remain nearly constant at low $T$ instead of decaying to zero.
There are several possible explanations for such a behaviour, one of them being a further magnetic phase transition for $T<0.4$~K (below the lowest experimentally attainable temperature). 
It is also possible that the low-$T$ part of the heat capacity of \sno\ is affected by a nuclear magnetism contribution.

Although to fully address an important question about the magnetic entropy involved in the observed transition in \sno\ the heat capacity measurements need to be extended down to the mK range, we show in Fig.~\ref{Fig_HC}(c) the area below the $C_{\rm mag}/T$ curve integrated in two different ways.
Solid symbols represent a straightforward integration from $T=0.4$~K upward, while the open symbols are obtained by extending the $C_{\rm mag}/T$ linearly down to 0 at $T=0$~K.
The visible difference between the two methods used seen in Fig.~\ref{Fig_HC}(c) helps to appreciate the importance of the low-$T$ part of the heat capacity measurements.
Regardless of the detail of the integration, in the first approximation \sno\ could be considered as an effective spin-1/2 system for which the amount of magnetic entropy recovered is $R \ln 2$.
Only approximately half of this entropy is recovered below the $T_{\rm N}$.

\begin{figure}[tb] 
\includegraphics[width=0.81\columnwidth]{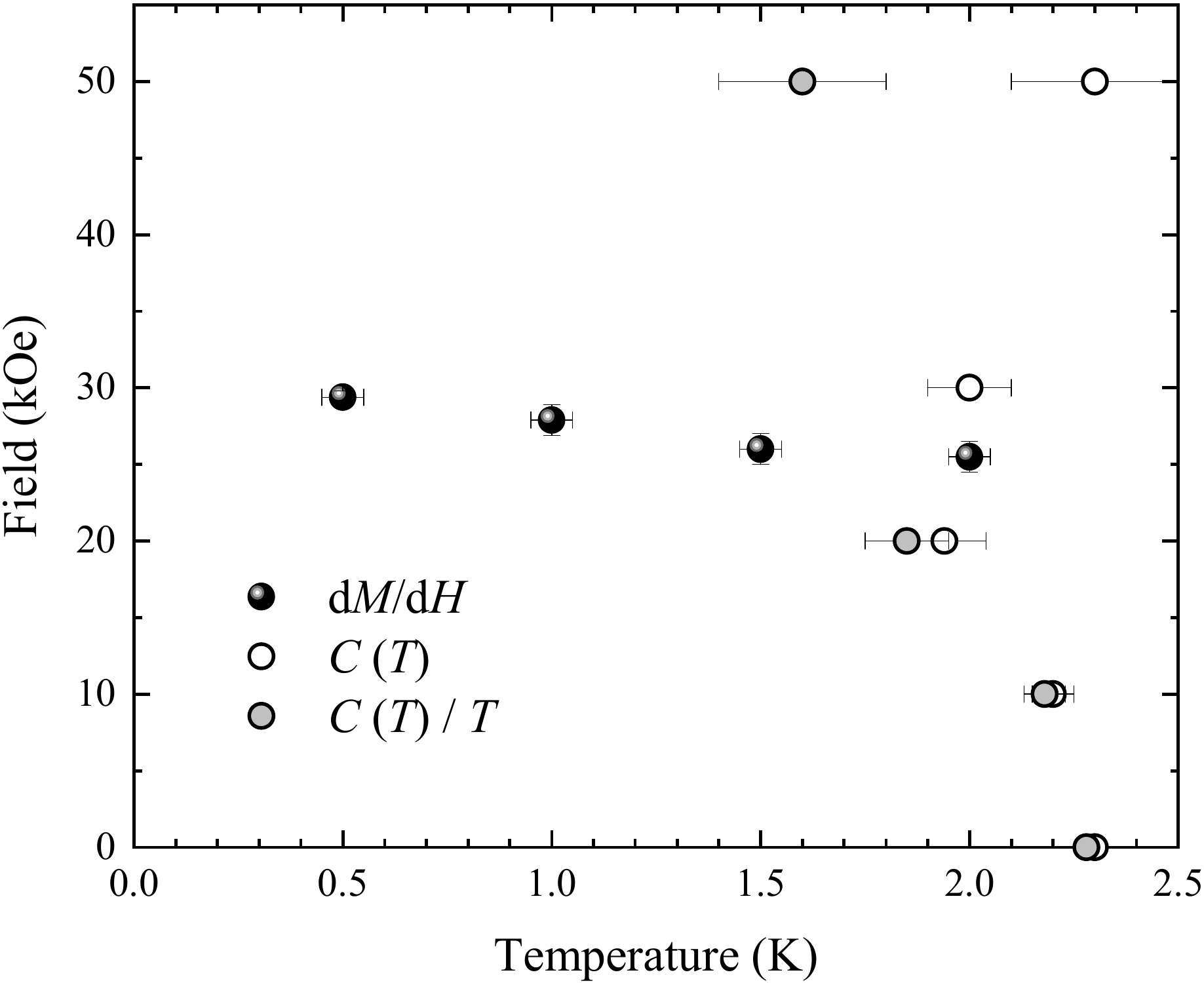}
\caption{Tentative magnetic phase diagram of \sno\ as determined from the magnetisation and heat capacity measurements.
	The black symbols correspond to the maxima in the $dM(H)/dH$ curves while the open and grey symbols mark to the maxima in the $C(T)$ and $C(T)/T$ curves respectively.}
\label{Fig_PhD}
\end{figure}

On application of a magnetic field [see Fig.~\ref{Fig_HC}(d)] the peak in $C(T)$ initially shifts to lower temperatures and becomes significantly broader.
It is still visible in applied fields of 30 and 40~kOe, however, the $C(T)/T$ curve in these fields does not show a well-defined maximum.
In the highest applied field of 50~kOe the maxima in both $C(T)$ and $C(T)/T$ are again clearly visible, although they are much broader and less defined compared to the low-field data.  
	
The positions of the maxima observed in the magnetisation $dM(H)/dH$ and heat capacity $C(T)$ and $C(T)/T$ curves are summarised in Fig.~\ref{Fig_PhD}.
The maxima are suggestive of the changes in magnetic state, however, one has to be rather careful in associating the features observed through the bulk properties measurements on a powder sample with positions of any phase transitions.
Nevertheless, combined magnetisation and calorimetry measurements suggest that even for the powder sample the ordering temperature in \sno\ is rather field sensitive, while the position of the field-induced transition only marginally changes with temperature.

\subsection{Inelastic neutron scattering}
\label{section_INS}

\begin{figure}[tb] 
\includegraphics[width=0.81\columnwidth]{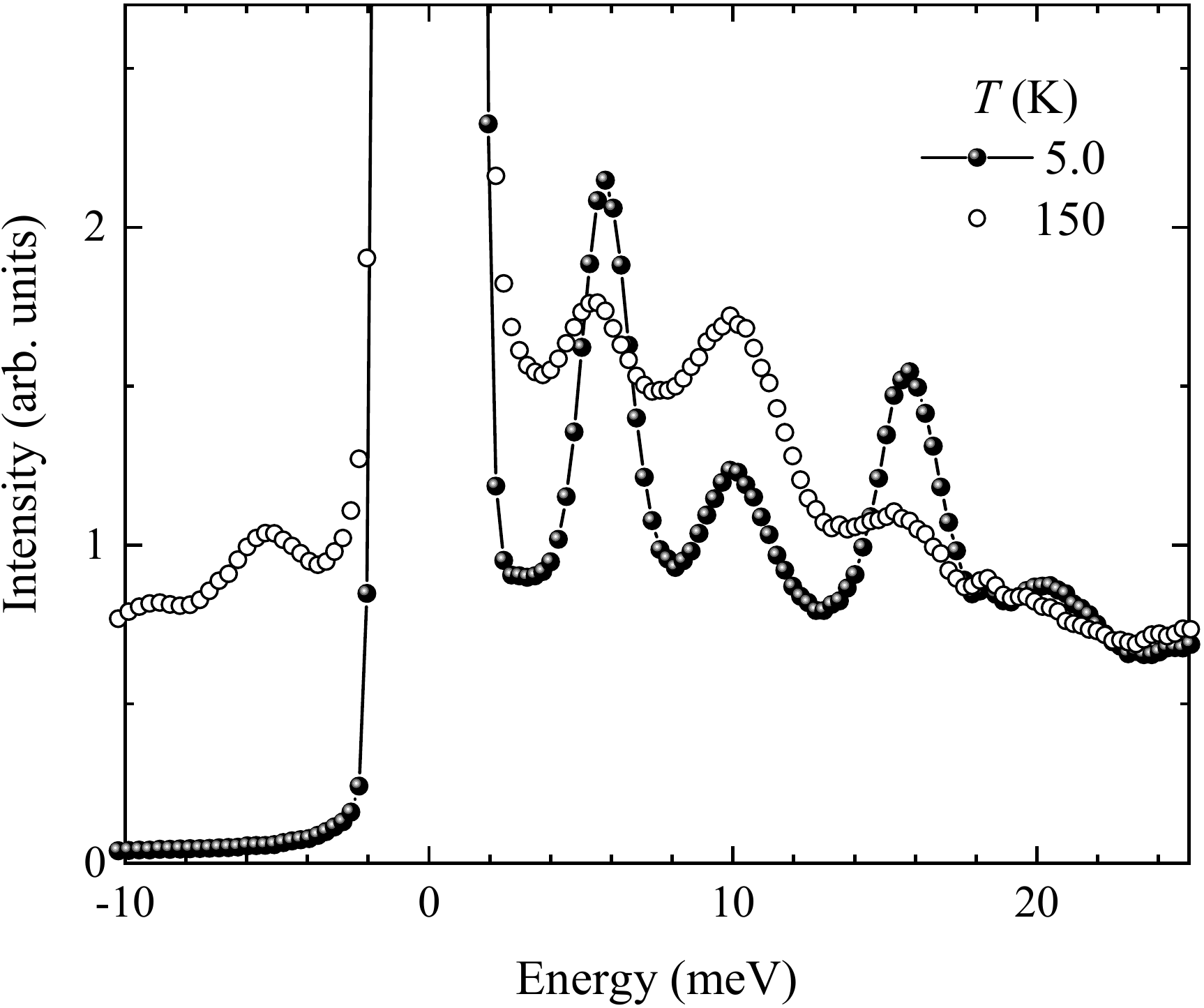}
\caption{Inelastic neutron scattering spectra of \sno\ at $T=5$~K (solid symbols) and $T=150$~K (open symbols) for wave vectors in the range $Q=(3 \pm 1)$~\AA$^{-1}$ measured with an incoming energy of 35~meV.
			Statistical errors are smaller than the symbol size.
			The line is a guide to the eye.}
\label{Fig_CEF}
\end{figure}

Preliminary results of the INS measurements are shown in Fig.~\ref{Fig_CEF}.
At a temperature of 5~K, there are two well-defined non-dispersive excitations at energies of 5.8 and 15.8~meV, their wave vector dependence (not shown) suggests that they are likely to be crystal field levels.
The data also show that the excitation at 10~meV has a different temperature dependence, which indicates that it might have a significant phonon contribution.
There is also a weak peak at 20.4~meV that might be of magnetic origin. 

The energy scale of the observed CEF excitations is similar to what has been reported for other members of the family~\cite{Fennell_2014,Malkin_2015}.
Because of the significantly different degree of distortions of the oxygen octahedra surrounding the two rare-earth sites for the various members of the \slo\ family, one could assume that as is seen in \seo, \sdo\ and \sho, the lowest CEF level at 5.8~meV is associated with only one rare-earth site in \sno.
It is therefore possible that only one Nd site should be included into the calculations of the CEF level contributions to the heat capacity [shown by the solid line in Fig.~\ref{Fig_HC}(a)].
However, a definitive contribution can only be calculated when the degeneracy of the CEF levels is established.
	
\subsection{Powder neutron diffraction}
\label{section_NPD}

The Rietveld refinement analysis of the data collected on the D2B diffractometer at room temperature confirmed the orthorhombic $Pnam$ space group of the crystal structure with all the ions occupying the $4c$ Wyckoff positions.
The refined crystal structure parameters of \sno\ at room temperature and at 5~K are shown in Table~\ref{TableI}. 

\begin{table}[tb] 
\caption{Refined structural parameters of \sno\ obtained from the PND experiments performed on the D2B instrument at 300 and 5.0~K respectively.
	All ions are in the Wyckoff $4c$ positions with coordinates (x, y, 0.25).}
\begin{ruledtabular}
\begin{tabular}{ccccc}
		& \multicolumn{2}{c}{300 K}		& \multicolumn{2}{c}{5.0 K}	\\
Atoms	&	x			&	y		&	x		& y			\\	\hline
Sr		&	0.7449(3)		&	0.6472(3)	&	0.7455(3)	&	0.6470(3)	\\
Nd1		&	0.4331(3)		&	0.1175(2)	&	0.4323(3)	&	0.1170(2)	\\
Nd2		&	0.4061(3)		&	0.6085(2)	&	0.4064(3)	&	0.6095(2)	\\
O1		&	0.7238(4)		&	0.3106(3)	&	0.7219(4)	&	0.3107(3)	\\
O2		&	0.6392(4)		&	0.0101(4)	&	0.6397(4)	&	0.0119(3)	\\
O3		&	0.4976(4)		&	0.7847(3)	&	0.4950(4)	&	0.7848(3) \\
O4		&	0.9345(5)		&	0.0824(3)	&	0.9322(4)	&	0.0826(3)	\\	\hline
$a$ (\AA)	& \multicolumn{2}{c}{10.152(1)}		& \multicolumn{2}{c}{10.128(1)} \\ 
$b$ (\AA)	& \multicolumn{2}{c}{12.205(1)}		& \multicolumn{2}{c}{12.177(1)} \\ 
$c$ (\AA)	& \multicolumn{2}{c}{3.5707(4)}		& \multicolumn{2}{c}{3.5639(3)} \\
\end{tabular}
\end{ruledtabular}
\label{TableI}
\end{table}

\begin{figure}[tb] 
\includegraphics[width=0.9\columnwidth]{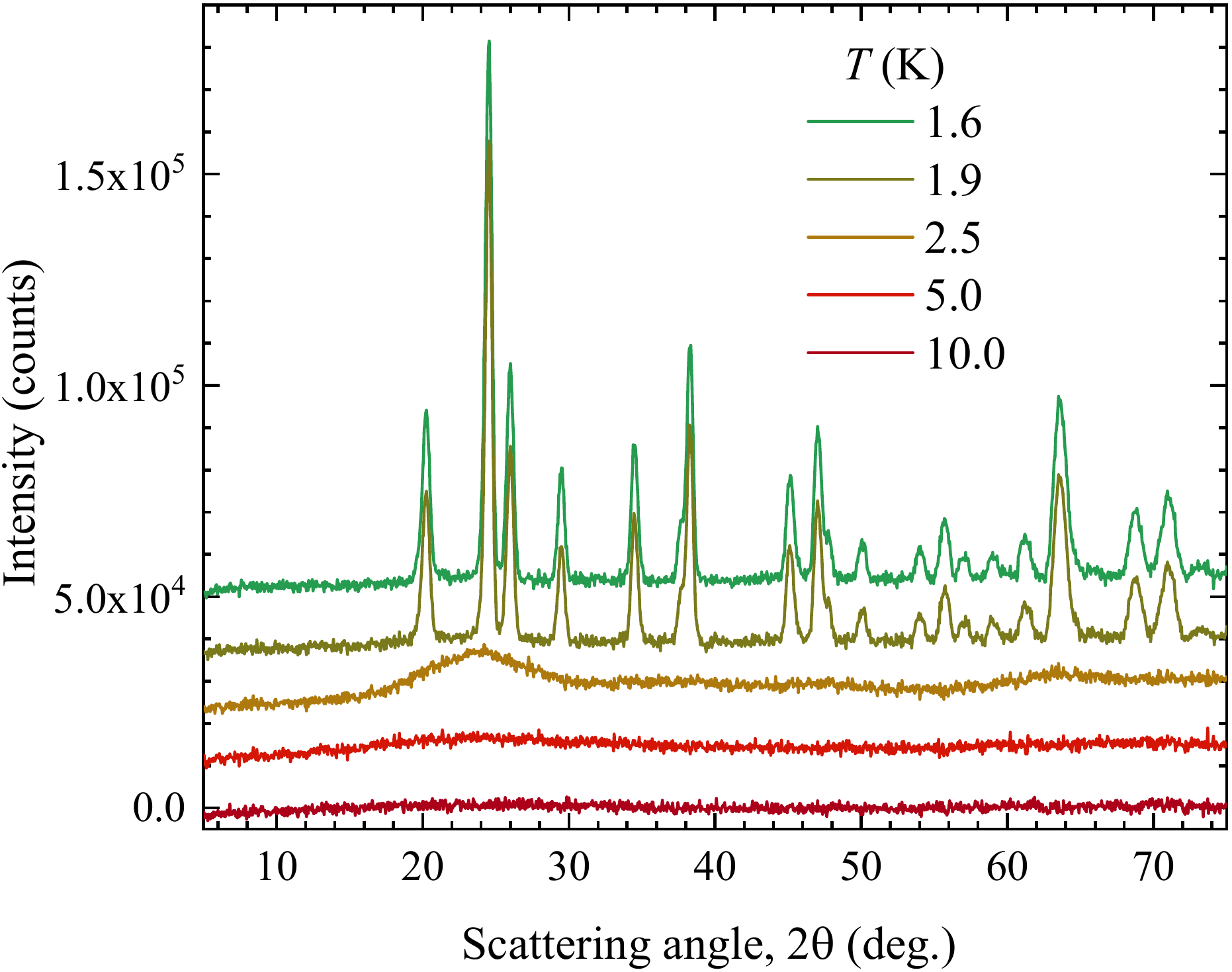}
	\caption{Temperature evolution of the magnetic neutron diffraction signal of \sno\ measured on the D20 powder diffractometer with $\lambda = 2.41$~\AA.
	The magnetic component is isolated from the total signal by subtracting a 25~K background.
	The patterns are consecutively offset by 15000 counts for clarity.}
\label{D20_varT}
\end{figure}

Magnetic PND patterns of \sno\ taken at various temperatures on the D20 diffractometer are shown in Fig.~\ref{D20_varT}.
The magnetic signal is isolated from the raw data by a subtraction of a higher-temperature background.
At the highest temperature shown, $T=10$~K, short-range magnetic order is barely visible as a very broad low-intensity feature at around $2\theta = 24^\circ$.
On cooling to 5~K the diffuse scattering is marginally more intense, however, when approaching the ordering temperature (see the $T=2.5$~K curve) it becomes much more structured and develops well-defined maxima at $2\theta = 24^\circ$ and $64^\circ$.
At this temperature, the observed diffuse scattering pattern (and therefore the structure of short-range magnetic correlations) is reminiscent of what has previously been observed at temperatures slightly above ordering in similar compounds such as \seo~\cite{Petrenko_2008}, \sho~\cite{Young_2012}, \bno~\cite{Aczel_2014}, \bto~\cite{Aczel_2015} and \bdo~\cite{Khalyavin_2019}.

On cooling below $T_{\rm N}$ the diffuse scattering is replaced by a collection of sharp magnetic Bragg peaks (as seen on the 1.9 and 1.6~K data sets in Fig.~\ref{D20_varT}) confirming the stabilisation of a long-range magnetic order.
However, a considerable amount of diffuse signal remains present in the data below $T_{\rm N}$ suggesting a retention of a significant disorder by the system.

A Rietveld analysis of the D20 diffraction data collected at 1.6~K was carried out after subtraction of a higher-temperature background and fixing the nuclear structure parameters to those obtained from the D2B data analysis.
The propagation vector ${\mathbf q}=(0 \frac{1}{2} \frac{1}{2})$ was determined from the positions of several strong magnetic reflections with the help of the k-search program implemented in the FullProf Suite~\cite{FULLPROF}.
An irreducible representation (irrep) analysis of the magnetic symmetry was carried out using the program BasIreps~\cite{FULLPROF}.

The analysis returns two 2-dimensional irreps referred to as $\Gamma_1$ and $\Gamma_2$, each irrep contains 6 basis vectors represented by the parameters $u$, $v$, $w$, $p$, $q$ and $r$.
The superposition of the symmetry-adapted basis vectors is shown as the magnetic modes $\psi_1$ and $\psi_2$ - corresponding to the irrep $\Gamma_1$ and $\Gamma_2$, respectively -  in Table~\ref{table_ireps}.
Due to the $\nicefrac{\pi}{2}$ phase shift between the Nd atoms 1 and 2, as well as between the atoms 3 and 4, resulting from the 2-fold screw axis parallel to $z$, we have used the constraints $u=p$, $v=q$ and $w=r$ which lead to a constant amplitude on each of the 4 positions.
Note that the same configuration can be achieved with $p=q=r=0$ and a global phase shift of $\nicefrac{\pi}{4}$, however, the refined parameters $u$, $v$ and $w$ need to be divided by $\sqrt{2}$ in order to obtain the magnetic moment size.
In a first refinement step the $w=r$ component of the magnetic moment proved to be insignificant, suggesting magnetic moments in the $a$-$b$ plane, and was set to 0 in the following.
	
\begin{table}
\caption{\label{table_ireps}
	Magnetic modes $\psi_n$ ($n = 1, 2$) of the two 2-dimensional irreducible representations $\Gamma_n$ for the Nd ions $m$ = 1-4 at given fractional coordinates ($x$, $y$, $z$) associated with a propagation vector $\mathbf{q} = (0 \frac{1}{2} \frac{1}{2}$).
	The basis vector components $u$, $v$, $w$, $p$, $q$ and $r$ connected to the spin $S_{\Gamma_n}^{m}$ have been refined respecting their symmetry constraints.}
	\begin{ruledtabular}
	\begin{tabular}{cccc}
	Atom $m$		& 			Position							& $\psi_1$ & $\psi_2$  \\ \hline \\
		1		& $\begin{pmatrix} x \\ y \\ \nicefrac{1}{4} \end{pmatrix}$	& $\begin{pmatrix} u \\ v \\ w \end{pmatrix}$  + $i \begin{pmatrix} \bar{p} \\ \bar{q} \\ r \end{pmatrix}$  & $\begin{pmatrix} u \\ v \\w \end{pmatrix}$ + $\begin{pmatrix} \bar{p} \\ \bar{q} \\ r\end{pmatrix}$  \\ \\
		2		&  $\begin{pmatrix} \bar{x} \\ \bar{y} \\ \nicefrac{3}{4} \end{pmatrix}$ & $\begin{pmatrix} \bar{p} \\ \bar{q} \\ \bar{r} \end{pmatrix}$  + $i\begin{pmatrix} u \\ v \\ \bar{w} \end{pmatrix}$  & $\begin{pmatrix} \bar{p} \\ \bar{q} \\ \bar{r} \end{pmatrix}$  + $i\begin{pmatrix} u \\ v \\ \bar{w} \end{pmatrix}$   \\ \\
		3		& $\begin{pmatrix} x+\nicefrac{1}{2} \\ \bar{y}+\nicefrac{1}{2} \\ \nicefrac{1}{4} \end{pmatrix}$  & $\begin{pmatrix} u \\ \bar{v} \\ \bar{w} \end{pmatrix}$  + $i \begin{pmatrix} \bar{p} \\ q \\ \bar{r} \end{pmatrix}$  & $\begin{pmatrix} \bar{u} \\ v \\ w \end{pmatrix} $  + $i\begin{pmatrix} p \\ \bar{q} \\ r \end{pmatrix}$  \\ \\
		4		& $\begin{pmatrix} \bar{x}+\nicefrac{1}{2} \\ y +\nicefrac{1}{2} \\ \nicefrac{3}{4} \end{pmatrix} $  & $\begin{pmatrix} p \\ \bar{q} \\ \bar{r} \end{pmatrix}$  + $i \begin{pmatrix} \bar{u} \\ v \\ \bar{w}\end{pmatrix} $  & $\begin{pmatrix} \bar{p} \\ q \\ r \end{pmatrix}$  + $i \begin{pmatrix} u \\ \bar{v} \\ w\end{pmatrix} $ 
	\end{tabular}
	\end{ruledtabular}
\end{table}

\begin{figure}[tb] 
\includegraphics[width=0.9\columnwidth]{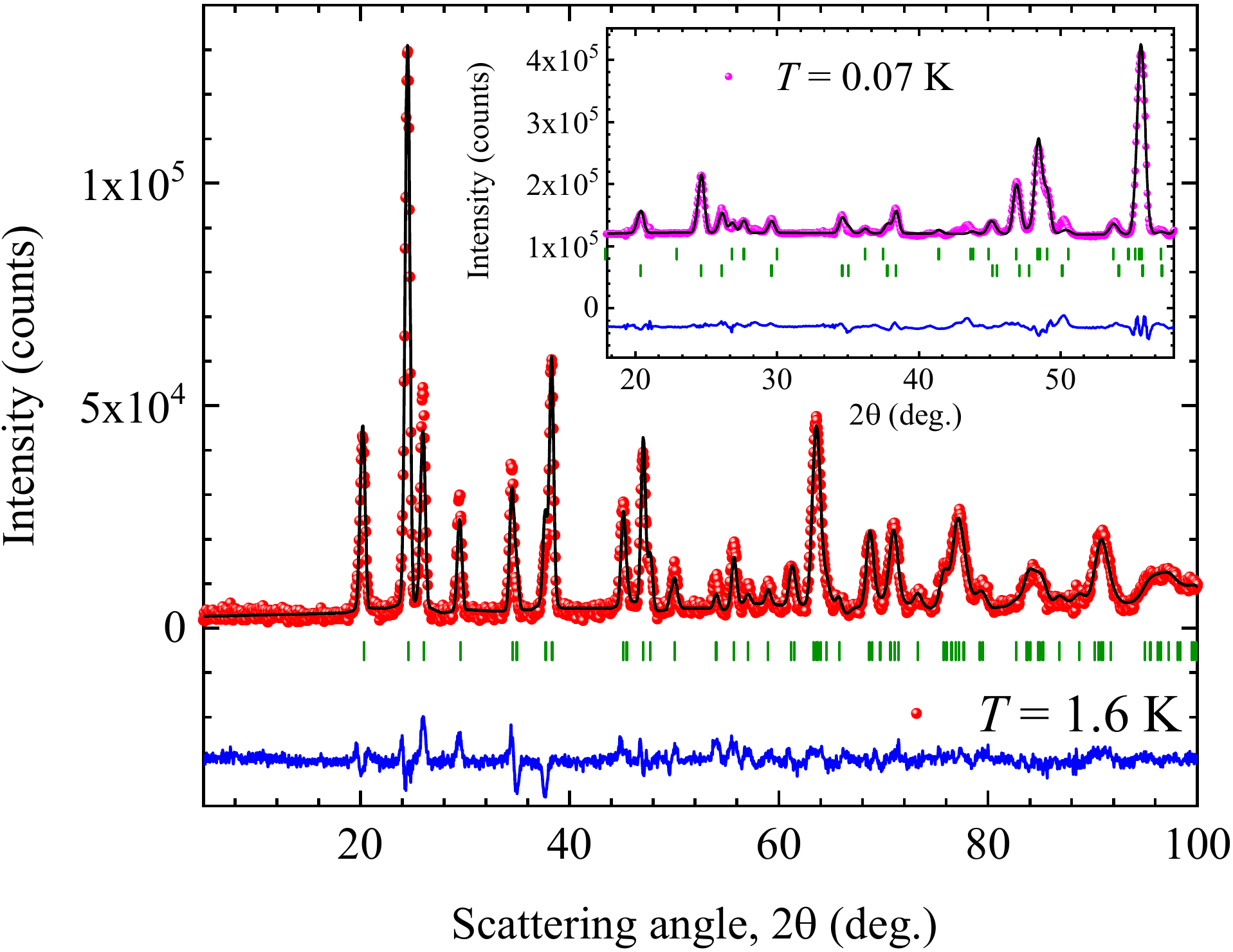}
\caption{Refinements of the magnetic diffraction pattern collected at 1.6~K on the D20 diffractometer and isolated from the raw data by subtracting a 10~K background.
	Data, fit and their difference are displayed in red, black and blue, respectively.
	The positions of the ${\mathbf q} = (0 \frac{1}{2} \frac{1}{2})$ symmetry allowed reflections are marked by vertical green bars.
	The inset shows the lower scattering angle section of the raw data recorded at $T=0.07$~K with a dilution refrigerator and a Cu sample can.
	The refinement at this temperature was performed on the total data, the vertical bars indicate the allowed nuclear (upper row) and magnetic (lower row) reflections.}
\label{V_can}
\end{figure}

By consecutively employing the two models to refine the magnetic data, we report for both cases an excellent agreement obtained between the data and fit ($R_F = 7.2$ and 7.0 for $\Gamma_{1}$ and $\Gamma_{2}$, respectively).
The observed and calculated patterns are shown in Fig.~\ref{V_can}.

 \begin{figure}[b]
	 \includegraphics[width=0.45\textwidth]{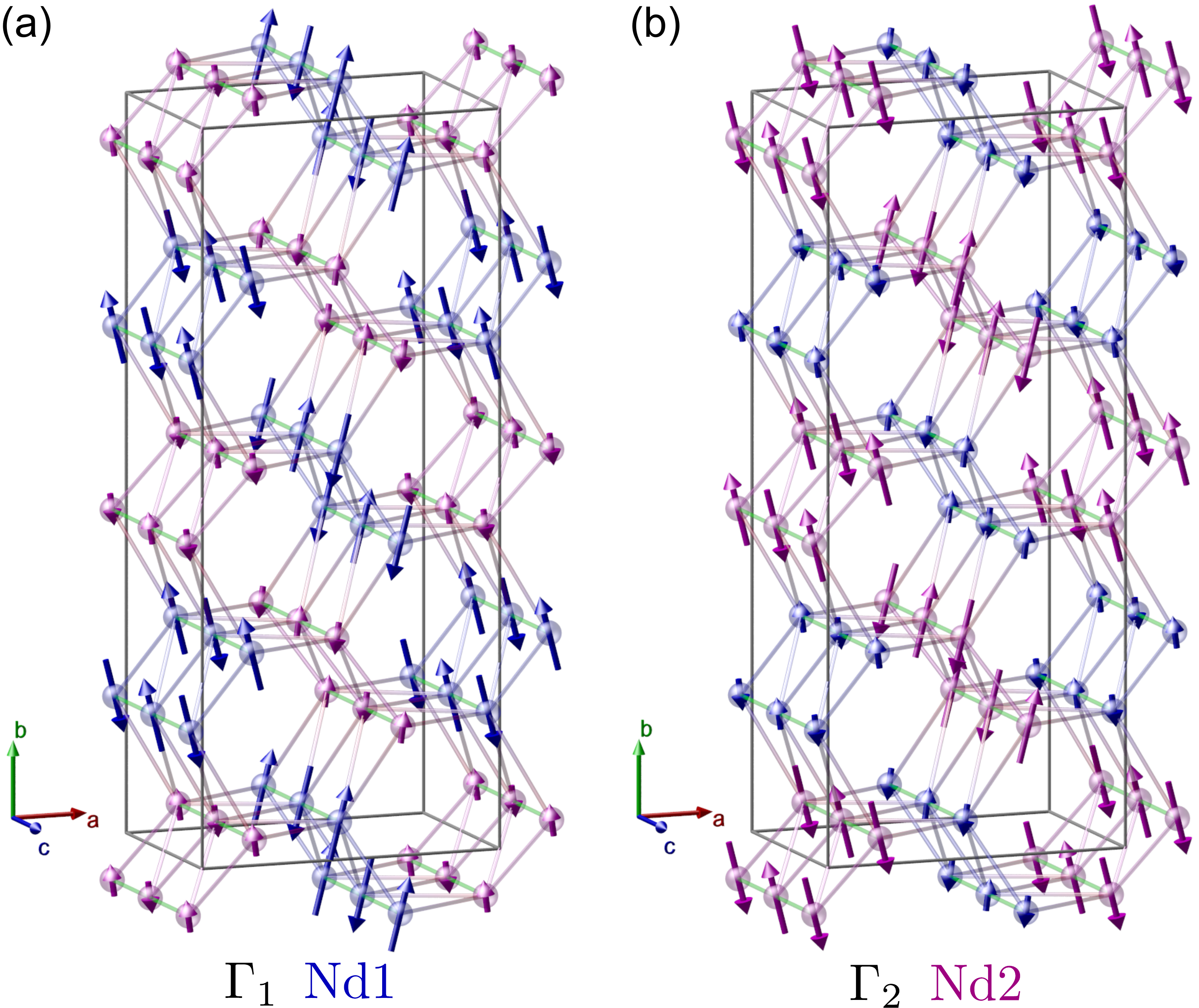}
	 \caption{Visualization of the magnetic structures stabilised by the \sno\ material below $T_{\rm N}$.
	The two structures were obtained by subsequently employing (a) the $\Gamma_1$ representation with the large magnetic moment on site Nd1 and (b) the $\Gamma_2$ representation with the large magnetic moment on site Nd2.}
	 \label{Magnetic_structure}
 \end{figure}

\begin{figure}[tb]
 	\includegraphics[width=0.45\textwidth]{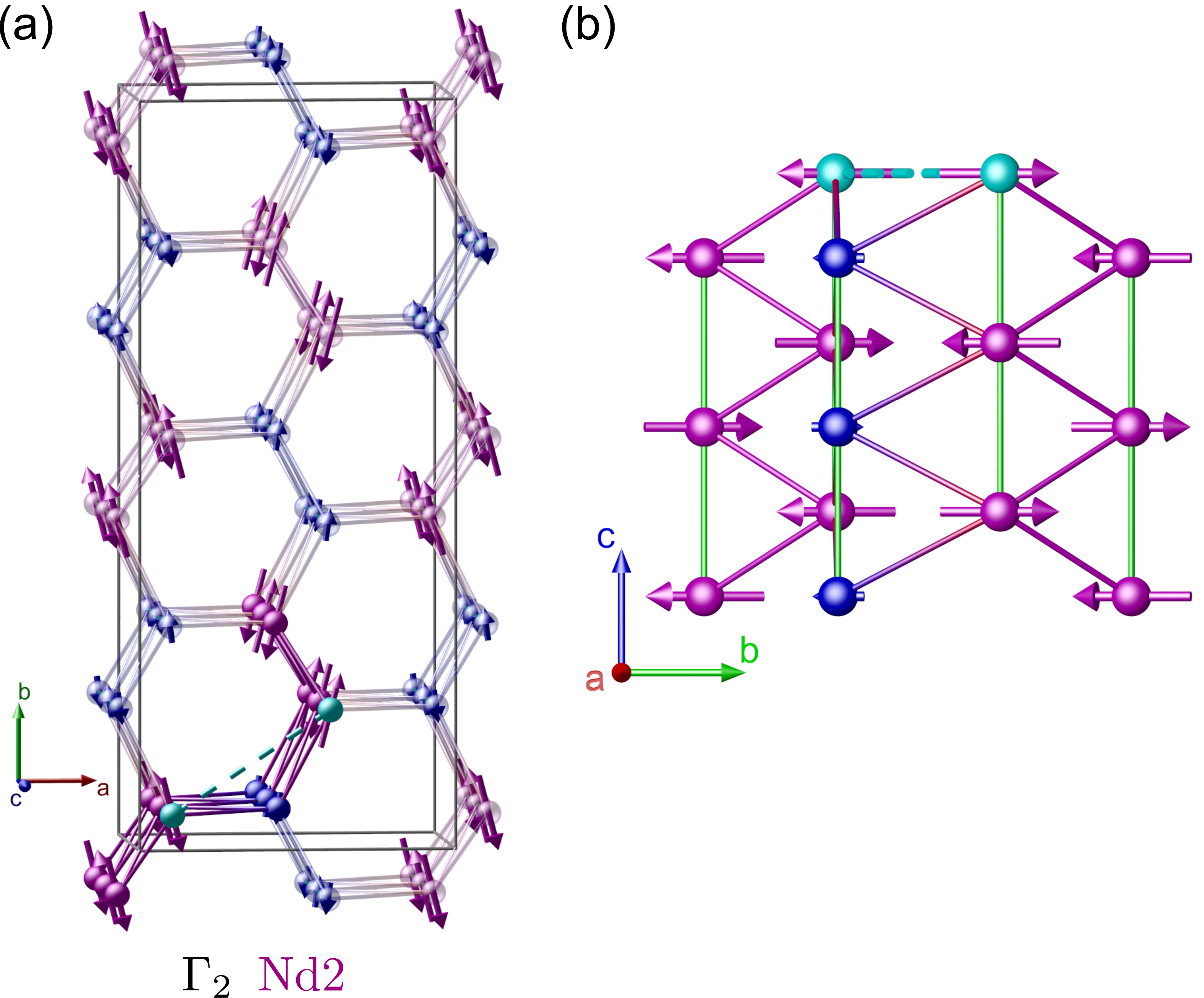}
 	\caption{Representation of the magnetic structure stabilised by the \sno\ material when viewed along the (a) $c$~axis and (b) $a$~axis.
	These perspectives allow for a clear visualisation of the inter-ladder antiferromagnetic ordering stabilised by the long-range ordered phase.
	The inter-ladder exchange interactions suspected to induce this ordering are displayed as cyan dashed lines.
	Note that that the section of the ladders shown in (b) corresponds to the non-transparent objects in (a).}
 	\label{irreps_J3}
 \end{figure}

The refined magnetic structures consist in both cases of a double N\'eel order where the magnetic moments lie within the crystallographic $a$-$b$ plane.
The refinement also shows that the long-range order is established on only one of the Nd sites.
The magnetic moments of the other Nd sites remain in a largely disordered state, indicating the presence of magnetic disorder below $T_{\rm N}$.
In fact, the results show that the \sno\ long-range ordered magnetic structure is obtained either by considering the $\Gamma_1$ symmetry with a mostly ordered magnetic moment on the site Nd1 or alternatively the $\Gamma_2$ symmetry with site Nd2 hosting the ordered magnetic moment, both options are shown in Fig.~\ref{Magnetic_structure}.
The two remaining combinations $\Gamma_1$-site~2 and $\Gamma_2$-site~1 return very poor agreement with the data and are therefore disregarded.
The difference between the two pairs of combinations lies in the symmetry of the same-site inter-chain spin ordering, see Fig.~\ref{irreps_J3}.
This particular characteristic can be easily appreciated when focusing on the next-nearest same-site neighbour within a hexagon revealing a primarily \afm\ inter-ladder alignment for the $\Gamma_1$-site~1 and $\Gamma_2$-site~2 configurations while $\Gamma_2$-site~1 and $\Gamma_1$-site~2 configurations (not shown here) would result in primarily ferromagnetic inter-ladder alignment.
Although the results of the PND refinements do not allow one to associate the magnetically ordered component with a particular Nd site, it is very likely that the ordered moment is on the Nd2 site.
As in many other members of the family, the larger distortions of the oxygen octahedra are around the second rare-earth site and these distortions are typically associated with the easy-plane type of magnetic anisotropy, as opposed to the Ising-type on the first site.
	 
Given such a distinct difference between the two pairs of possible spin arrangements ($\Gamma_1$-site~1 and $\Gamma_2$-site~2 on the one hand and $\Gamma_2$-site~1 and $\Gamma_1$-site~2 on the other hand), one has to question whether the \afm\ inter-ladder exchange interactions in \sno\ are indeed strong enough to make a selection.
The presence of the \afm\ inter-ladder interactions has previously been suggested in the studies of another member of the same family, \seo~\cite{Hayes_2011}.
As the distance between the Nd ions for this bond is rather long, just above 6~\AA, the influence of dipolar interactions is ruled out, however, one could imagine that the indirect exchange involving, for example, two oxygen ions, O1 and O3 in the notations of Table~\ref{TableI} is not negligibly small.
In this case the distances between the Nd and O atoms are 2.299 and 2.260~\AA, while the O-O distance is 3.097~\AA.

The results of the Rietveld refinement analysis carried out subsequently using the two irreducible representations returned in both cases a magnitude of about 2~$\mu_{\rm B}$ for the long-range ordered site while less than 0.3~$\mu_{\rm B}$ is obtained on the second site, which remains mainly disordered at $T=1.6$~K, well below $T_{\rm N}$, see Table~\ref{magnetic_moments}. 

The rather small ordered component established on the principally disordered site has been clearly evidenced by the Rietveld refinement analysis and hence, should not be neglected.
In fact, performing refinements of the data in the $\Gamma_1$-site~1 or $\Gamma_2$-site~2 configurations when fixing the value of the ordered moment to zero respectively on site~2 and site~1, increases the $R_F$ factor considerably.
Furthermore, a refinement that simultaneously includes the $\Gamma_1$ and $\Gamma_2$ models, on site~1 and 2, respectively, and also forces the same moment components on both sites returns a significantly higher $R_F$ factor value.
These results thus justify the use of the configurations involving a single irrep and resulting in a small ordered component in addition to the principally ordered phase.
As the small ordered component is either in the $\Gamma_1$-site~2 or $\Gamma_2$-site~1 configurations, it induces inter-ladder ferromagnetic ordering.
Consequently, a ferromagnetic inter-ladder interaction could play a notable role in this system. 
	
\begin{table}[tb]
\caption{Components of the magnetic moments on the two Nd sites of \sno\ determined through the refinements of the 1.6 and 0.07~K data.
The long-range ordered phase is attributed to the Nd2 site considering the larger distortions of the oxygen octahedra as observed in other members of the \slo\ family.}
\centering
\begin{ruledtabular}
\begin{tabular}{lllllll}
	&\multicolumn{2}{c}{$\mu_a$ ($\mu_{\rm B}$)}	&\multicolumn{2}{c}{$\mu_b$ ($\mu_{\rm B}$)} 	& \multicolumn{2}{c}{$\mu_{tot}$ ($\mu_{\rm B}$)}		\\ 
 	&	1.6 K		&	0.07 K				&	1.6 K		&	0.07 K				&	1.6 K		&	0.07 K	\\ \hline
Nd1	&	0.04(4) & 0.1(1) & 0.29(2) & 0.28(7) & 0.29(2) & 0.30(8) \\ 
Nd2	&	0.52(3) & 0.5(1) & 1.98(1) & 1.89(4) & 2.05(2) & 1.96(5) \\
\end{tabular}
\end{ruledtabular}

\label{magnetic_moments}
\end{table}

In order to track potential low temperature magnetic phase transitions that could lead to the long-range ordering of the remaining disordered Nd sites, we have collected scattering patterns at 1.0, 0.7, 0.3 and 0.07~K, on the D20 instrument.
From this analysis, we report a clear stability of the long-range ordered magnetic phase down to the lowest temperature (the raw data at $T = 0.07$~K are displayed in the inset of Fig.~\ref{V_can}).
The patterns at 0.07 K and 1.6 K do not show any significant differences, from which we conclude the temperature stability of the double N\'eel  magnetic phase between $T_{\rm N}$ and the lowest temperatures.

To appreciate the temperature evolution of the magnetic moments, a Rietveld refinement analysis was also performed on the raw data collected at 0.07~K.
This analysis was executed with a simultaneous refinement of the \sno\ nuclear and magnetic phases, in addition to the Cu phase (sample container) responsible for the presence of intense Bragg reflections seen in the diffraction pattern.
A good agreement was obtained between data and fit for both $\Gamma_{1}$-site~1 and $\Gamma_{2}$-site~2 models, confirming the temperature stability of the magnetic phase down to 0.07~K, see Table~\ref{magnetic_moments}.
The weak moments of the second type of sites still remain largely disordered even at 0.07~K.

\begin{figure}[tb]
\includegraphics[width=0.87\columnwidth]{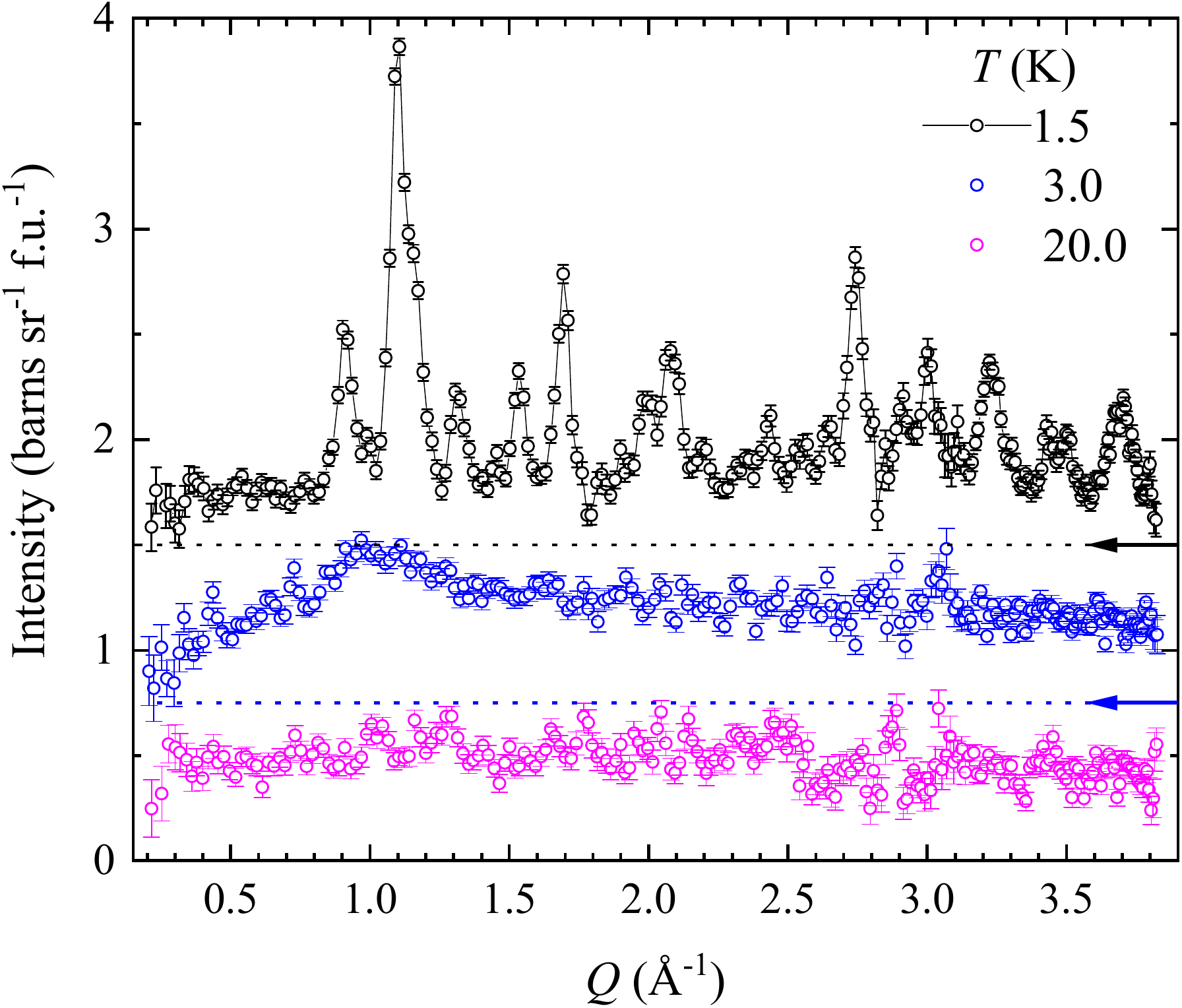}
	\caption{Temperature evolution of the magnetic scattering signal from a powder sample of \sno\ collected on the D7 polarisation analysis instrument.
	The data for 1.5 and 3.0~K are offset for clarity with the zero for each pattern indicated by a colour-coded arrow.}
	\label{D7}
\end{figure}

The presence of a magnetically disordered component in \sno\ at temperatures well below $T_{\rm N}$ is also seen in the diffraction data collected on the D7 diffractometer.
The polarisation analysis technique allows for a clear identification of the magnetic signal without the need for higher-temperature background subtraction.
The magnetic signal measured on D7 at 20, 3.0 and 1.5~K is shown in Fig.~\ref{D7}, and it reveals the presence of a significant diffuse scattering component at all three temperatures studied.

At $T=20$~K, the diffuse magnetic signal shows almost no $Q$-dependence reflecting on a largely paramagnetic behaviour of \sno\ at this high temperature, while on approaching $T_{\rm N}$, at $T=3$~K, a sizeable diffuse feature is stabilised at low-$Q$.
This feature is somewhat characteristic of the \slo\ compounds and has been seen in several of them~\cite{Karunadasa_2005} indicating a buildup of strong short-range magnetic correlations in the system above the $T_{\rm N}$.
The measurements performed below the ordering transition, at 1.5~K, reveal a significant reduction of the diffuse scattering component compared to the 3.0 and 20~K observations, however, the diffuse scattering signal remains easily visible over a wide $Q$ range at this temperature.
In this respect the behaviour is similar to what has been reported for \bno~\cite{Aczel_2014}, a Nd containing compound which orders magnetically with the same wave vector as \sno.

An interesting observation from the D7 results is that even at the higher temperature, $T=20$~K, the magnetic signal does not simply decay with $Q$ as expected due to the magnetic form-factor.
One possible reason for this is the contribution from the inelastic scattering, as the measurements on D7 are energy integrated.
For the wavelength used, 3.14~\AA, the incident energy is 8.3~meV and the integration will inevitably include the lowest-energy CEF level at 5.8~meV, which will make the $Q$ dependence of the intensity less trivial.
An example of a similar effect and detailed explanations of the the kinematic conditions are given in a recent publication~\cite{Kimber_2020}.
	
\section{Conclusions}
The low temperature magnetic properties of \sno\ were studied using powder neutron scattering techniques as well as inelastic neutron scattering, magnetisation and specific heat measurements.
The appearance of a long-range \QV\ \afm\ double N\'eel order below $T_{\rm N}=(2.28\pm0.04)$~K is established through the Rietveld refinement analysis of neutron diffraction data.
The refined magnetic structure possesses a sizeable ordered magnetic moment (confined largely within the $a$-$b$ plane) for only one of the two Nd$^{3+}$ sites.
The magnetic moments on the other Nd$^{3+}$ site have only a small ordered component and remain principally disordered down to the lowest temperatures. 
	    
The magnetisation and heat capacity measurements indicate the presence of a field-induced transition at around 30~kOe.
We suggest that this is a reorientation type of the transition, while the saturation takes place in much higher fields. 		

\section*{ACKNOWLEDGMENTS}
	
We are grateful to B.Z.~Malkin a for numerous discussions of the magnetic properties of various \slo\ compounds and also to M.R.~Lees for careful reading of the manuscript.
	
 \bibliography{SrLn2O4_all}
\end{document}